\documentclass{aa}

\usepackage{txfonts}
\usepackage{graphicx}
\usepackage{amssymb}
\usepackage{color}

\author{Antoine M\'erand\inst{1} \and 
  Pascal Bord\'e\inst{2}\fnmsep\thanks{Michelson Postdoctoral Fellow,
    formerly at LESIA, Observatoire de Paris.} \and 
  Vincent Coud\'e du Foresto\inst{1}}

\authorrunning{M\'erand et al.}

\title{A catalog of bright calibrator stars for 200-meter baseline \\
  near-infrared stellar interferometry\thanks{The catalog is available
  in electronic form at the CDS via anonymous ftp to
  cdsarc.u-strasbg.fr (130.79.128.5) or via
  http://cdsweb.u-strasbg.fr/cgi-bin/qcat?J/A+A/(vol)/(page)}}

\titlerunning{A catalog of bright calibrators for Interferometry}

\institute{LESIA, UMR8109, Observatoire de Paris, 5 place Jules
   Janssen, 92195 Meudon, France \\
   \email{Antoine.Merand@obspm.fr, Vincent.Foresto@obspm.fr}
   \and
   Harvard-Smithsonian Center for Astrophysics, 60 Garden Street,
   Cambridge, MA 02138, USA \\
   \email{pborde@cfa.harvard.edu}}

\date{Received --- / Accepted ---}

\abstract{ 
We present in this paper a catalog of reference stars suitable for
calibrating infrared interferometric observations. In the K band,
visibilities can be calibrated with a precision of 1\,\% on baselines
up to 200~meters for the whole sky, and up to 300~meters for some part
of the sky.  This work, extending to longer baselines a previous
catalog compiled by Bord\'e et al. (2002), is particularly well
adapted to hectometric-class interferometers such as the Very Large
Telescope Interferometer (VLTI, \cite{vlti}) or the CHARA array
(\cite{chara}) when observing well resolved, high surface brightness
objects ($K \la 8$). We use the absolute spectro-photometric
calibration method introduced by Cohen et al. (1999) to derive the
angular diameters of our new set of 948 G8--M0 calibrator stars
extracted from IRAS, 2MASS and MSX catalogs. Angular stellar diameters
range from 0.6\,mas to 1.8\,mas (median is 1.1\,mas) with a median
precision of 1.35\,\%. For both the northern and southern hemispheres,
the closest calibrator star is always less than $10\degr$ away.


}
  
\begin{document}

\maketitle
%
%
\section{Introduction}
Long baseline stellar interferometers measure the amount of coherence
(ie the squared modulus of the coherence factor, sometimes also
called the raw visibility amplitude, or raw fringe contrast) between
any pair of their subapertures. This quantity needs to be calibrated
in order to yield the true visibility of the source, which is the
modulus of the Fourier transform of the object's intensity
distribution at the spatial frequency $B/\lambda$ defined by the
projected baseline $B$ and the wavelength $\lambda$. The squared
visibility $V^2$ of an object is derived from the measure $\mu^2$ of
its coherence factor, and from an \textit{interferometric efficiency}
factor $\mathcal{T}^2$ (also called \textit{transfer function}),
which accounts for the coherence losses caused by imperfections of the
instrument and by the Earth's turbulent atmosphere:
\begin{equation}
V^2 = \frac{\mu^2}{\mathcal{T}^2}.
\label{V2T2}
\end{equation}
From Eq.~\ref{V2T2}, it appears that the errors on ${\mathcal{T}^2}$
and ${\mu^2}$ contribute equally to the accuracy of the object squared
visibility $V^2$. As the interferometric efficiency varies during the
night owing to changing instrumental and atmospheric conditions, it
has to be frequently sampled, and then interpolated for the time of
the science target observation (one can see for example
\cite{Perrin98} for details of an interpolation strategy) to ensure a
reliable calibration.The $\mathcal{T}^2$ determination is done by
observing a reference source ($\mu_\mathrm{cal.}^2$) with a known
squared visibility ($V_\mathrm{cal.}^2$), for which

\begin{equation}
\mathcal{T}^2 \equiv \frac{\mu_\mathrm{cal.}^2}{V_\mathrm{cal.}^2}.
\label{trans}
\end{equation}

In practice, observations of the science target are interleaved with
observations of reference stars, referred to as \textit{calibrator
stars} or \textit{calibrators} for short. The choice of a calibrator
is critical as one should be able to predict its intrinsic visibility
at the projected baseline of the interferometer, with an accuracy
which should be at least as good as the measurement of the raw
visibility (more about this in Section~\ref{sec:choice}). For this
purpose, \cite{Borde} carefully extracted a catalog of calibrator
stars from the spectro-photometric reference stellar network compiled
by Cohen et al. (1999), hereafter \cite{CohenX}. Because of the
estimated angular diameter of the selected stars (typically 2.3~mas
with an error of 1.2~\%), this catalog, hereafter Cat.~1, provides
calibrators for interferometric baselines up to $\sim 100$~m in the
near infrared. The advent of large interferometers, such as the VLTI
and CHARA arrays with maximum baselines of 202 and 330~m,
respectively, motivated us to compile a catalog for longer baselines,
hereafter Cat.~2.

In Sect.~\ref{sec:choice}, we explain why the sensitivity of current
interferometers leads us to take as calibrators partly resolved stars
instead of point-like sources, in the bright star regime. Then, we
argue that these stars can be safely modeled as uniform disks provided
they are carefully selected, and we review two different strategies to
make this selection. Last, we explain how to decide if a calibrator is
appropriate for a given observation. In Sect.~\ref{sec:building}, we
describe how we built Cat.~2 by using \cite{CohenX}'s method and
infrared photometric measurements for stars fainter than in
Cat.~1. Finally, in Sect.~\ref{sec:charac}, we discuss the main
characteristics of Cat.~2. In particular we demonstrate the relevance
of this new catalog for the VLTI/VISA (the VLT interferometer
sub-array, involving the 1.8\,m auxiliary telescopes) and the CHARA
array.

%
%
\section{Choosing and modeling calibrators} \label{sec:choice}


The ideal calibrator is a source for which the intrinsic visibility,
using the same observational setup as the science target, can be
perfectly predicted. Thus every calibrator needs to be described by a
morphological model, so that its true squared visibility can be
predicted for the full range of spatial frequencies that can be
addressed by the interferometer. As reference sources are chosen among
stellar objects, the most common models employed to describe a
calibrator are, in increasing number of free parameters:

\begin{enumerate}
\item The point source (no free parameters): in that case
  $V^2_\mathrm{cal.}=1$ at any baseline;
\item The uniform disk (UD model), with the angular UD diameter $\theta$
  as the only free parameter: the monochromatic squared visibility is
  then given by the relationship:
  \begin{equation} \label{eq:UD}
    V_\mathrm{UD}^2(x) = 4\left(\frac{J_1(x\theta)}{x\theta} \right)^2,
  \end{equation}
  where $J_n$ denotes the nth Bessel function of the first kind and x
  the spatial frequency, namely $\pi B/\lambda$ (B, baseline and
  $\lambda$ the observing wavelength);
\item
  The limb darkened disk (LD model), with two or more free parameters.
\end{enumerate}

For simplicity and reliability, more complex calibrator models should
be avoided. These include non centro-symmetric or variable
morphologies like pulsating or flare stars, fast rotators, and binary
stars. Indeed, a good calibrator needs to be:

\begin{enumerate}
\item {\em Well modeled}, in the sense that the model which is used to
describe the morphology of the object is as simple as possible, yet
appropriate within the level of accuracy required for the visibility;
\item {\em Well known}: the free parameter(s) in the model need to be
known with sufficient accuracy;
\item {\em Stable}: no time dependency of the morphology;
\item {\em Observable with the same setup} and in the same conditions as the
science target, e.g. close to the target on the sky, with comparable
magnitude and preferably spectral type.
\end{enumerate}


It should be noted that these requirements are conflicting since in
general they call for references much smaller than the science
targets, yet having approximately the same color and apparent
brightness, which is impossible for thermal sources. Therefore the
choice of a calibrator will always be the result of a compromise. The
first and last requirements exclude the point source model in almost
every circumstance for long baselines, as it would imply an
unrealistically small (hence faint) reference. For example, in order
to induce a bias smaller than 1\,\% ($V^2>99\,\%$), a reference star
considered as a point source with a 300\,m baseline at
$\lambda=2.2\,\mu$m needs to have an angular diameter smaller than
0.1\,mas. If the source is a K0 giant, this corresponds to an infrared
magnitude $K \ga 8$, beyond the sensitivity limit of interferometers
with small or medium size telescopes (less than 2\,m in diameter)
without fringe tracking devices.  For interferometers using larger
telescopes, with a sensitivity limit close to or better than $K=10$,
it appears possible to find real non-resolved sources ($V^2>99\%$),
because sensitivity is no longer an issue. Still, one has to be
careful for the following reasons. A $K=10$ star, earlier than M5,
will definitely be unresolved for a 200-m baseline interferometer
operating in the near infrared, but one has to ensure that this
particular candidate follows the chosen model, i.e. a point
source. Following our previous discussion, the {\it well known} item
is out of concern for an unresolved star, in the sense that no
parameter drives the model, but the {\it well modeled} one still
applies. For example, the candidate may be binary, just like the
majority of main sequence stars (eg see \cite{Quist2000}).  Therefore,
non-sensitivity limited interferometers still need well modeled (thus
trusty) calibrators. Considering that faint candidates ($K\ga 8$) will
arise from infrared photometric surveys, such as DENIS or 2MASS, for
which very few supplementary data are available, the potential
binarity will be very difficult to detect. It appears that there are
two different cases for modern hectometric baseline stellar
interferometers: the bright star regime ($K \la 8$) and the faint star
regime ($K \ga 8$). Because one wants the interferometer to operate in
the same regime for both the scientific target and the calibrator (eg
for signal to noise ratio issue), we advocate that an extension to the
\cite{Borde} catalog, containing only well modeled (uniform disk
ensured) and well known (high accuracy estimation of the angular
diameter), will not only be useful for medium aperture arrays
operating in the near infrared, such as the CHARA Array or the Palomar
Testbed Interferometer (\cite{PTI}), but also for the VLTI/VISA
interferometer using AMBER in the bright star regime.

\begin{figure}
   \begin{center}
     \resizebox{\hsize}{!}{\includegraphics{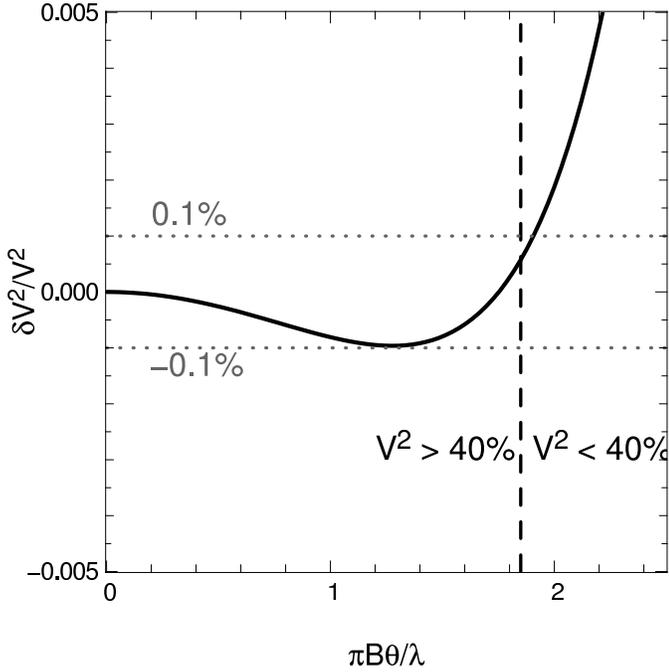}}
     \caption{Relative difference between UD and LD squared
     visibilities ($\delta V^2/V^2$) with respect to spatial frequency
     times angular diameter. The UD diameter is adjusted to minimize the
     quadratic deviation between UD and LD squared visibilities
     in the $V^2>40\,\%$ domain.
     This particular LD model represents the most limb-darkened disks of
     our catalog, namely for M0III stars in the J band. In the domain
     corresponding to $V^2>40\,\%$, the relative difference remains under
     0.1\,\%. It is even smaller for less limb-darkened stars.}
     \label{justif}
   \end{center}
\end{figure}

It happens that quiet single stars can easily and correctly be modeled
as uniform disks when the squared visibility exceeds $\approx 40~\%$,
as the difference between a UD and a LD model is then smaller than 0.1\,\%
(see Fig.~\ref{justif}). Stellar angular diameters can be known from direct
high-angular resolution measurements (interferometry or Lunar
occultations) or estimated indirectly by (spectro-)photometry. We
argue that a spectrophotometric estimate of the diameter is more
suitable for reference purposes, because direct diameter measurements
have only been performed on a limited number of sources
(listed in the Catalog of High Angular Resolution Measurements
(\cite{charm}), hereafter CHARM), and most of these sources do not have
the properties required for being a good calibrator at long
baselines. Besides, this interferometric data set is in essence
heterogeneous and more specifically, UD values are published usually
at a single wavelength. Because of limb darkening, these values are not
readily valid for other band-passes. Therefore, in absence of an
exhaustive and homogeneous set of direct diameter measurements of
suitable calibrator stars (a certainly much needed, however intensive,
observational program), we prefer in the remainder of this paper to
apply an indirect method to determine the diameters for a set of stars
that are selected so that they can be trusted as single and stable.

%
At this point, two strategies have been proposed:
\begin{enumerate}
 \item \emph{A dynamic or open list approach}: a software that queries
existing databases for calibrator candidates in a given field is
designed and distributed.  The software is expected to help the
observer in the down-selection process by using the information
available in the databases. Then, it should provide either a direct
estimate of the visibility at the time of the observation, or the
parameter value(s) for a given visibility model (eg the UD angular
diameter). Two examples of this approach are \emph{Search Calibrator}
integrated in the software
ASPRO\footnote{http://mariotti.ujf-grenoble.fr/$\sim$aspro/} issued by
the Jean-Marie Mariotti Center (\cite{Bonneau04}), and
\emph{getCal}\footnote{http://msc.caltech.edu/software/getCal/} issued
by the Michelson Science Center;
\item \emph{A static or closed list approach}: a catalog of carefully
selected potential calibrators and including all the necessary
information to compute the visibility for any relevant baseline and
bandpass, is compiled and distributed. Such a list, the \emph{ESO
Calvin Tool}\footnote{http://www.eso.org/observing/etc/preview.html},
has been made available for the VLTI Mid-Infrared interferometer
(MIDI, see \cite{midi}).
\end{enumerate}

We prefer the closed list approach for three reasons:
\begin{enumerate}
\item It enables a dedicated check on all of its entries, which would
  not be possible with a generic selection process;
\item The catalog can be revised and updated as some of its entries
  are found -- either serendipitously or by means of a systematic
  interferometric survey -- to be unreliable as a reference (the most
  likely reason being that they are hitherto unknown binaries), with
  the objective that eventually all bad entries will be removed;
\item Selecting a calibrator is easier for the beginner in
  interferometry since it can be hand-picked from a source list.
\end{enumerate}

However, the dynamic approach has its merits as it could
provide either specific calibrators missing in our catalog (eg stars
earlier than K0), or standard calibrators for the faint star regime
($K \ga 8$) where a closed list is unmanageable.

\section{The calibrator decision diagram}
The purpose of the calibrator decision diagram described here is to
find the right calibrator for a given observation. In order to achieve
the desired precision for a given instrumental configuration,
calibrators have also to be selected according to their diameters and
diameter errors. For the UD model, the stellar angular diameter
contains all the knowledge of the source, thus the squared visibility
error associated with the calibrator error is
\begin{equation}
\left( \frac{\sigma_{V^2_{\mathrm{cal.}}}}{V^2_{\mathrm{cal.}}}
\right)_\theta = \frac{\partial
V^2_{\mathrm{cal.}}/\partial\theta}{V^2_{\mathrm{cal.}}}= 2x\theta
\frac{J_2(x\theta)}{J_1(x\theta)} \frac{\sigma_\theta}{\theta}.
\label{VarV}
\end{equation}

As the knowledge of the calibrator should not be the limiting factor
on the precision of the visibility measurement, we require for the
calibrator choice that the relative error on the squared visibility,
$\sigma_{V^2_{\mathrm{cal.}}}/V^2_{\mathrm{cal.}}$, due to the
uncertainty on the calibrator diameter alone, $\sigma_\theta/\theta$,
should not exceed the internal (instrumental) error,
$\sigma_{\mu^2}/\mu^2$, that is to say
\begin{equation}
\left(\frac{\sigma_{V^2_{\mathrm{cal.}}}}{V^2_{\mathrm{cal.}}}\right)_{\theta}
\le \left(\frac{\sigma_{\mu^2}}{\mu^2}\right)_\mathrm{limit}.
\label{decide}
\end{equation}

For further discussions, we set the upper limit on inequality
(\ref{decide}) to 2\,\% as it implies on the relative error on the
visibility an upper limit of $\sigma_V/V = 1\,\%$, the standard
precision benchmark for single-mode interferometers in the
literature. Examples of scientific results achieved with such a
precision include the study of the internal structure of Sirius by
\cite{alphacenVINCI} or the test of the model atmosphere of a M4 giant
by \cite{Wittkowski2004}.

Inequality~\ref{decide} can be turned into a \textit{decision diagram}
that helps decide whether a calibrator is suitable or not for a given
observation. In the framework of the UD model and leaving aside
magnitude and spectral type considerations, a given source can be
represented as a point in a diameter/diameter error plane, ie
$(\theta,\,\sigma_\theta/\theta)$. We refer to this plane as the
\textit{calibrator plane}. As for the instrumental configuration, all
needed information is embedded in the pair spatial
frequency/instrumental precision, i.e. $(x,\,(\sigma_{\mu^2} /
\mu^2)_\mathrm{limit})$ or alternatively in the triplet
wavelength/projected baseline/instrumental precision, ie
$(\lambda,\,B,\,(\sigma_{\mu^2} / \mu^2)_\mathrm{limit})$. By holding
two of these three quantities fixed, one can plot the third one as a
function of the diameter and diameter error. We call this plot the
\textit{calibrator decision diagram}.


\begin{figure}
   \begin{center}
     \resizebox{\hsize}{!}{\includegraphics{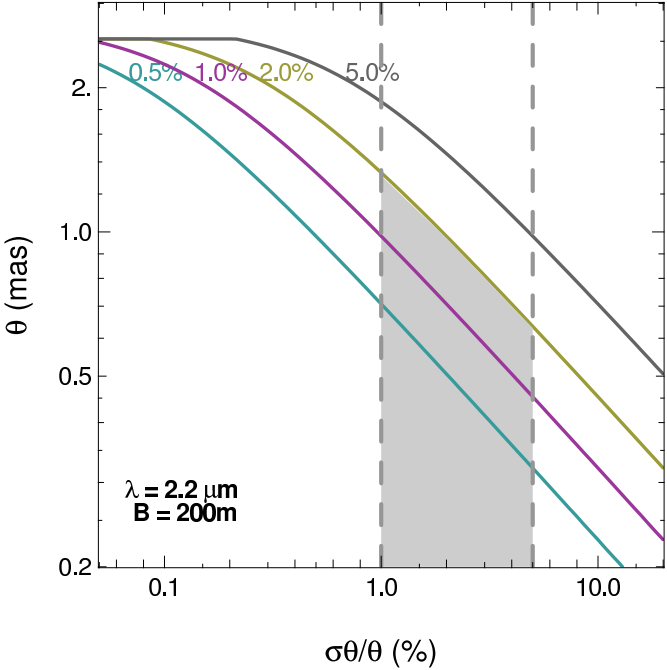}}
     \caption{Calibrator decision diagram (logarithmic scales) fixing
       the wavelength to $2.2\:\mu\mathrm{m}$ and the baseline to
       200~m: contours of the instrumental precision as a function of
       the diameter and diameter error of the calibrator. Stars below
       a given contour could be used at the labeled
       precision. Vertical dashed lines delimit the area corresponding
       to the typical diameter error obtained through
       (spectro-)photometric determinations (1--5~\%).}
     \label{explain}
   \end{center}
\end{figure}

As a first example, we set on Fig.~\ref{explain} the wavelength to
$2.2\:\mu\mathrm{m}$, the baseline to 200~m, and we plot contours of
the instrumental precision in the calibrator plane. In that case, if
one wants $\sigma_{V^2_{\mathrm{cal.}}}/V^2_{\mathrm{cal.}}$ to be at
most 2~\%, the diagram shows that either $\theta \le
1.3\:\mathrm{mas}$ with a 1~\% error or $\theta \le
0.65\:\mathrm{mas}$ with a 5~\% error is required. It is noteworthy
that for $\sigma_{V^2_{\mathrm{cal.}}}/V^2_{\mathrm{cal.}} \le 2\:\%$
and $\sigma_{\theta}/\theta \ge 1\:\%$, the squared visibility stays
above 40~\% (corresponding to a $\theta=1.3\:\mathrm{mas}$ star,
observed at $\lambda=2.2\:\mu$m, with a 200\,m baseline), which
assures of the validity of the UD model as discussed previously. For
this reason, we keep in the following these requirements on the
squared visibility and angular diameter precisions.

{\color{red}  }


\begin{figure}
   \begin{center}
     \resizebox{\hsize}{!}{\includegraphics{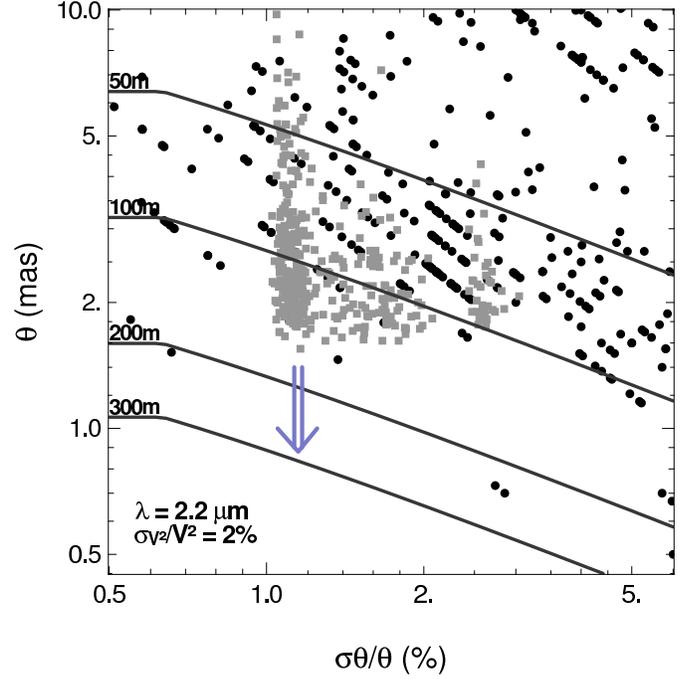}}
     \caption{Calibrator decision diagram (logarithmic scales) fixing
     the wavelength to $2.2\:\mu\mathrm{m}$ and the instrumental
     precision to 2~\%: contours of the baseline as a function of the
     diameter and diameter error of the calibrator. Stars below a
     given contour can be used at the labeled baseline. Stars from
     Cat.~1 appear as gray squares and stars from CHARM as black
     dots. The double arrow indicates the direction we chose to
     follow to expand Cat.~1.  The 50, 100, 200 and 300\,m curves in
     the K band ($\lambda=2.2\,\mu m$) corresponds respectively to 35,
     75, 145 and 220\,m in the H band ($\lambda=1.6\,\mu m$) and 25,
     55, 110 and 165\,m in the J band ($\lambda=1.2\,\mu m$).}
     \label{CatBorde}
   \end{center}
\end{figure}

As a second example, we set in Fig.~\ref{CatBorde} the wavelength to
$2.2\:\mu\mathrm{m}$, the instrumental precision to 2~\%, and we plot
contours of the baseline in the calibrator plane. Black dots
represent stars flagged as single and with no shell in CHARM, and gray
squares represent stars from Cat.~1. Stars located below a given
contour can be used as calibrators at the corresponding baseline. It
appears from this plot that CHARM provides calibrators for projected
baselines up to $\sim 70\:\mathrm{m}$, and Cat.~1 for baselines up to
$\sim 100\:\mathrm{m}$, not being too stringent on the sky coverage.


%
%
\section{Building the catalog} \label{sec:building}
\subsection{How to expand Cat.~1: better models or smaller stars?}
Looking at Fig.~\ref{CatBorde}, there are two ways in which Cat.~1
could be expanded to satisfy the 2~\% precision at 200~m baselines:
\begin{enumerate}
\item Decrease the diameter errors by improving the calibrator models
   (going leftwards on the diagram);
\item Decrease the stellar diameters by finding smaller stars (going
   downwards on the diagram).
\end{enumerate}

The first solution necessitates working with more resolved stars which
necessitates more elaborate models than the UD, along with
high-quality observations to determine the parameters of these models.
Keeping with our objectives of simplicity and reliability, we favor
the second approach, which consists in going only to smaller stars since
we guaranty hereby the validity of the UD model. On the other hand, we have
to be very careful not to degrade the error. The second solution
features the drawback of working with fainter stars. Indeed, the
angular diameter logarithmic scale on Fig.~\ref{CatBorde}, can be
viewed as a linear magnitude scale for a given spectral type (a given
temperature).

\subsection{Pushing the spectro-photometric method further} \label{crit5}
Through their whole series of papers Cohen et al. built an all-sky
network of absolute reference sources for spectro-photometry in the
infrared. The tenth paper, \cite{CohenX}, exposes at length their
method: by fitting low-resolution composite spectra to infrared
photometric measurements, they derive both absolute spectral
irradiances and stellar angular diameters. Their grid stars were
selected from the IRAS Point Source Catalog (hereafter IRAS~PSC) by
applying stringent quality criteria that qualify most of these stars
as interferometric calibrators as well. Cat.~1 was then built
by removing the very few stars that would depart from a UD at low
resolution.

In their original work, \cite{CohenX} used the following 6 criteria
in order to select calibrator candidates in the IRAS~PSC:
\begin{enumerate}
\item The IRAS flux at $25\:\mu\mathrm{m}$ must be greater than 1~Jy;
\item IRAS infrared colors are restricted to one particular quadrant,
  corresponding to giant stars;
\item IRAS candidates must be as normal as possible: sources flagged as
  non-stellar, variable, emission line or carbon stars are rejected;
\item Total IRAS fluxes within a radius of $6\arcmin$ must be less than
  5~\% of the IRAS candidate corresponding fluxes;
\item IRAS stars must not be associated with a small extended
  source. The contribution of infrared cirrus must be less than 5\% of
  the candidates infrared flux;
\item To reduce the chance of selecting variables, candidates must
  fall in one of the following spectral and luminosity classes:
  \begin{itemize}
  \item A0--G9 and II--IV,
  \item K0--M0 and III--V.
  \end{itemize}
\end{enumerate}

Walking in the tracks of \cite{CohenX}, we applied the previous set of
criteria to the IRAS PSC, modifying criterion 1 to
${F_{25\:\mu\mathrm{m}} \le 1\:\mathrm{Jy}}$ in order to sort out new
stars. We also required the stars to be identified in the SIMBAD
database to make possible further quality filtering (criteria 3 and
6). A total of 8313 stars passed criteria 1 and 2, as well as the
identification by SIMBAD. In the selection process, we found the 5th
criterion -- meant to remove stars blended with interstellar cirrus --
severely impairs the sky coverage around the Galactic Equator.
Therefore, we decided to leave aside that criterion, and to rely
instead on the quality of the spectrum fit to the infrared photometric
measurements.

\subsection{Discarding candidates that depart from a UD}
Multiplicity and variability are the main causes of departure from a
UD that one would like to avoid. Ideally, multiple stars should be
rejected according to the separation and contrast between the
components. As thresholds on these quantities could only be determined
according to the instrument configuration and characteristics
(baseline, wavelength, accuracy, etc.), we chose to remove all known
spectroscopic and astrometric binaries on the basis of their SIMBAD
object type. However, because of SIMBAD's hierarchical classification,
a star can have a companion without being classified as a binary. For
this reason, we also used the Catalog of Visual and Double Stars in
Hipparcos (\cite{Dommanget00}) to remove all stars with a companion
within $2\arcmin$. This distance represents a safe upper limit for any
pointing devices or tip-tilt servos.  In addition, the Hipparcos
Catalog (\cite{hip}) was used in order to remove known variables, as
well as stars with a positive ``Proxy'' flag, ie having field stars
within $10\arcsec$. The $2\arcmin$ and $10\arcsec$ thresholds make the
catalog useful for a wide variety of instruments without impacting
significantly on the number of stars left in the catalog (only 12\,\%
of the stars were discarded).
 
In order to pin down more variable stars or unresolved binaries, we
used several radial velocity catalogs, starting with the catalog by
\cite{Malaroda01} who collected all radial velocity measurements from
the literature until 1998. Every entry of this catalog contains the
nature of the object according to every author. If a single author
suspected a star to be a spectroscopic binary, we discarded that
candidate. Furthermore, we looked at \cite{deMedeiros99} and
\cite{Nidever02} for post-1998 measurements and used them in same
manner as \cite{Malaroda01}. This process led to remove an additional
1\,\% of our candidates.

A last cause of departure from the symmetric UD model would be
oblateness due to fast rotation. This should not be a concern in our
stellar sample as it contains only G8--M0 giants expected to be slow
rotators. We checked that by using $v \sin(i)$ as a proxy for the
rotation rate: indeed, for all 36 stars with measurements, we found
that $v \sin(i) < 10\,\mathrm{km/s}$.

As a general remark, let us point out that considering the large
number of candidate stars, it was out of question to check every one
of them in the literature, and to perform dedicated observations where
it would have appeared necessary. Therefore, we would like to draw the
attention of the reader to the fact that a few stars which are
improper for calibration may remain in our published list.

\subsection{Photometry}
Deriving angular diameters with \cite{CohenX}'s method requires
absolute photometric data between $1.2$ and $35\:\mu\mathrm{m}$. We
used $12\:\mu\mathrm{m}$ and $25\:\mu\mathrm{m}$ photometric data from
the IRAS~PSC (our input catalog), as well as those from the IRAS~FSC
(Faint Source Catalog). We extracted the near-infrared photometry, H
and Ks magnitudes, from the Two Micron All Sky Survey Point Source
Catalog (hereafter 2MASS~PSC) with the absolute calibration by
\cite{CohenXIV}. The 2MASS~PSC was the only near-infrared catalog that
matches our needs in terms of sky coverage. However, we had to deal
with a poor photometric precision ($\sigma_K = 0.25$) for the bright
stars ($K=3$--4) we are interested in, as they tended to saturate
2MASS detectors. Finally, the MSX infrared Astrometric Catalog
(\cite{Egan96}) provided 6 photometric data points in the
$4.22-4.36\:\mu\mathrm{m}$ (called B1 band),
$4.24-4.45\:\mu\mathrm{m}$ (called B2 band), $6.0-10.9\:\mu\mathrm{m}$
(called A band), $11.1-13.2\:\mu\mathrm{m}$ (called C band),
$13.5-16.9\:\mu\mathrm{m}$ (called D band) and
$18.1-26.0\:\mu\mathrm{m}$ (called E band) infrared bands. These last
measurements were converted into absolute photometric measurements
using \cite{CohenXII} calibration.

We looked for interstellar extinction by computing the spectral class
photometry (especially B-V), comparing it to its expected value. If
$A_\lambda$ is the extinction in magnitude, the interstellar medium is
characterized by $R_V = A_V/(A_B-A_V) = A_V/E_{B-V}$. We used an
average of $R_V=3.1$. Then, thanks to the tables of $A_\lambda/A_V$
from \cite{Cardelli89}, we corrected the H and K magnitudes for
reddening. However, this correction always stays within the
photometric error bars, because our candidates are bright stars that
are close and for which 2MASS photometry is not very
accurate. Therefore our relatively crude estimation of the reddening
is appropriate.

\subsection{Deriving limb-darkened diameters from stellar templates}
Stellar templates computed by \cite{CohenX} consist of absolute
composite spectra for different spectral types and luminosity classes,
along with limb-darkened (LD) angular diameters $\phi$. Because
\cite{CohenX} used plane-parallel Kurucz model atmospheres, their LD
angular diameters ($\phi$) correspond to Rosseland angular diameters,
associated with a Rosseland optical depth of unity. Indeed, this
diameter corresponds to the zone where photons from the continuum are
emitted, and where the limb brightness drops to zero. In the case of
plane-parallel models the Rosseland diameter is equal to the limb
darkened diameter $\phi$ (see discussion in \cite{Wittkowski2004}).

\begin{figure}
   \resizebox{\hsize}{!}{%
     \includegraphics{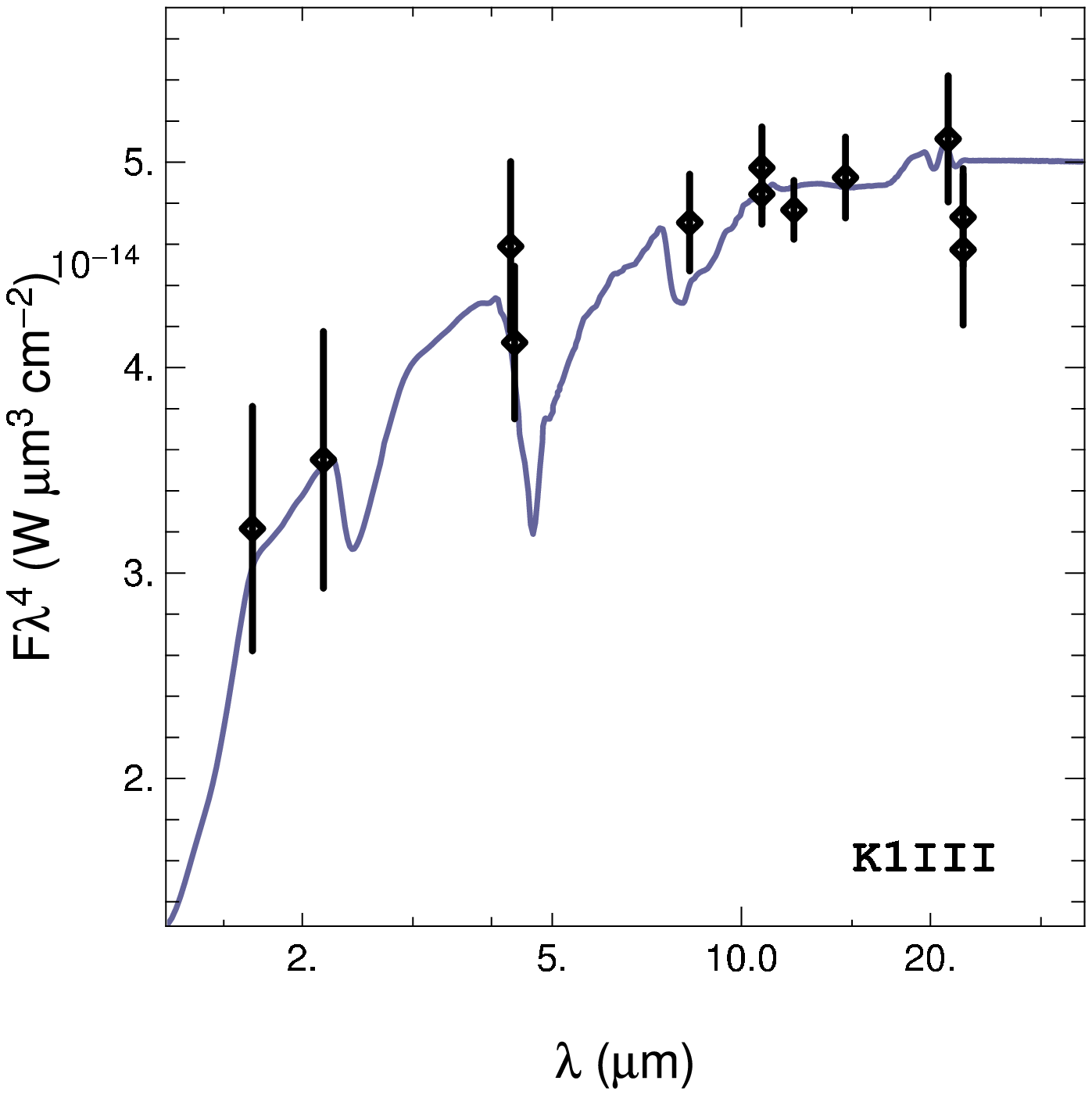}
     \includegraphics{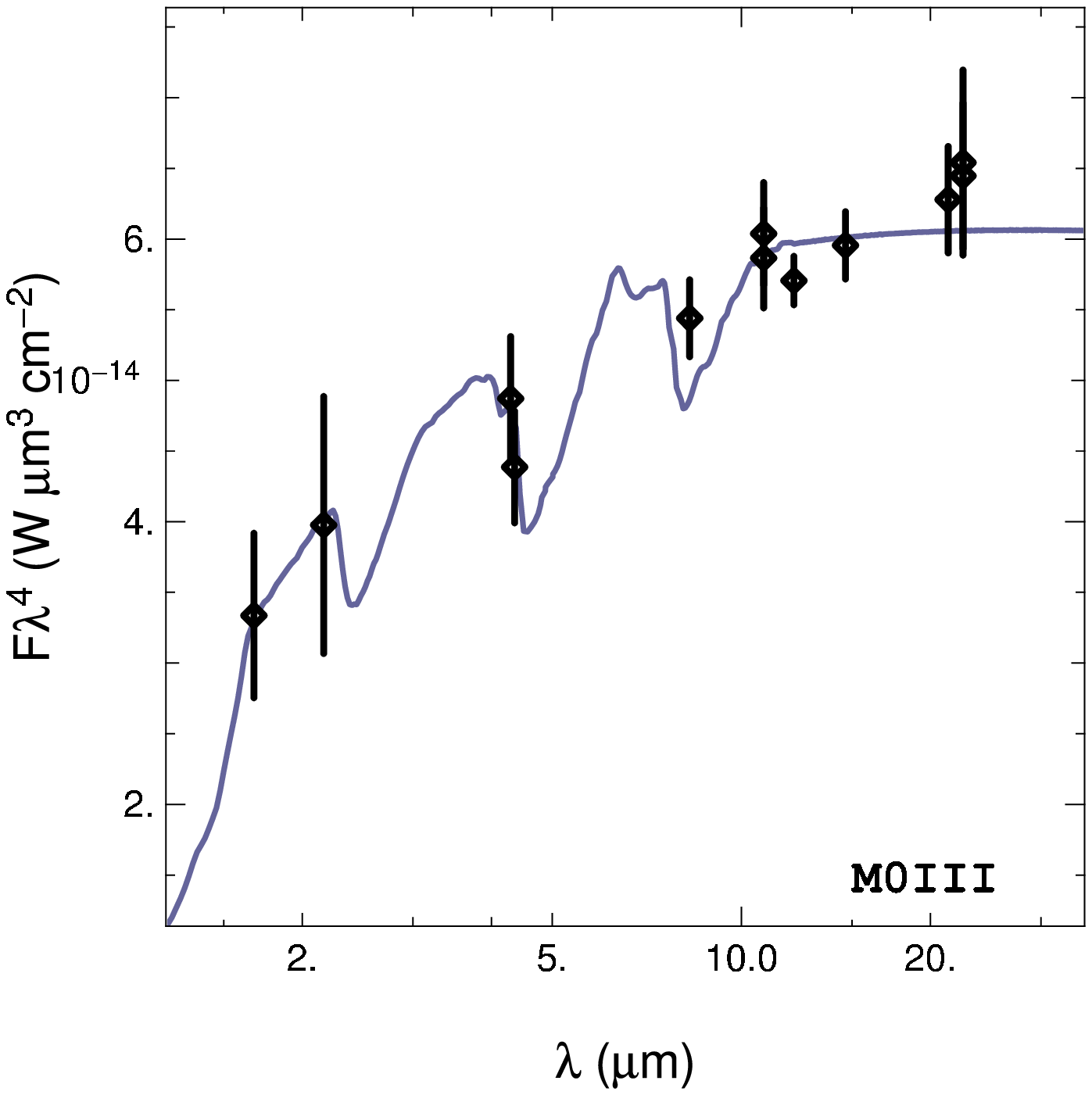}}
   \caption{Examples of stellar template fits to photometric
     measurements converted into monochromatic irradiances at
     isophotal wavelengths. The data points stand for (from left to
     right): H, $\mathrm{K_s}$ (2MASS), B1, B2, A (MSX), $F_{12\:\mu
     \mathrm{m}}$ (IRAS PSC and FSC), C, D, (MSX) and $F_{25\:\mu
     \mathrm{m}}$ (IRAS PSC and FSC).  \textit{Left:} HD~15248, a
     K1III star (${\chi^2_\mathrm{r} = 0.66}$).  \textit{Right:}
     HD~164724, a M0III star (${\chi^2_\mathrm{r} = 0.83}$).}
   \label{HDex}
\end{figure}

After retrieving the spectral types and luminosity classes of our
candidate calibrators from SIMBAD, we fitted the absolute template
corresponding to a specific star to its infrared photometric
measurements. All photometric measurements were converted into
isophotal quantities, following the calibration prescribed by C99 for
IRAS PSC and FSC, by \cite{CohenXIV} for 2MASS PSC and \cite{CohenXII}
for MSX. Fig.~\ref{HDex} shows two examples of stellar template fits.

The scaling factor resulting from the fit, $l$, yielded the LD diameter
of the star
\begin{equation}
\phi = \sqrt{l} \times \phi_\mathrm{ref}.
\end{equation}
For every diameter, we computed an associated error by taking into account the
contributions of the fitting formal error, $\sigma_l$, corresponding to
$|\chi^2(l)-\chi^2(l \pm \sigma_l)|=1$, and of the original template diameter
error, $\sigma_{\phi_\mathrm{ref}}$, according to
\begin{equation}
\left( \frac{\sigma_\phi}{\phi} \right)^2 =
\left( \frac{\sigma_l}{2l} \right)^2 +
\left( \frac{\sigma_{\phi_\mathrm{ref}}}{\phi_\mathrm{ref}} \right)^2.
\end{equation}

We could not find in \cite{CohenX} the angular diameter errors of
their original templates. By comparing the diameters published by
these authors and the ones computed by our implementation of their
method, this internal error was determined to be
$\sigma_{\phi_\mathrm{ref}}/ \phi_\mathrm{ref}= 1.03\%$. This value
appeared to be independent of spectral type and luminosity class.

As mentioned at the end of Sect.~\ref{crit5}, we used the quality of
the template fit as a filtering tool to ensure the consistency of our
method.  Namely, if the reduced $\chi^2$, i.e. $\chi^2_\mathrm{r} =
\chi^2/(N-P+1)$, where N is the number of data points and $P$ the
number of parameters, is greater than 1, we exclude the star from the
final selection.

\subsection{From limb-darkened to uniform disk diameters}
We converted LD diameters into UD diameters in J, H and K bands by using the
following four-coefficient limb-darkening law from \cite{Claret00}
\begin{equation} \label{eq:CLV}
I(\mu) = 1 - \sum_{k=1}^4 a_k \, (1-\mu^{k/2}),
\end{equation}
where $I$ is the specific intensity and $\mu$ the cosine of the angle
between the line of sight and the perpendicular to the stellar
surface.  \cite{Claret00} has tabulated the coefficients $a_k$
according to the stellar effective temperature $T_\mathrm{eff}$ and
surface gravity $\log (g)$.

Only 65 of our candidates could be found in the \cite{CdS} catalog of
effective temperature and surface gravity measurements, all of them
with a spectral type G8--K4. For all others, we estimated
$T_\mathrm{eff}$ by using the \cite{Bessell} polynomial relation based
on the color index $V-K$. When no measurements were available, we used
the typical surface gravity for the spectral class, as well as solar
abundances. This is legitimate since the infrared limb-darkening is
not very sensitive to these parameters.

Taking into account a center-to-limb variation, the visibility function
becomes
\begin{equation} \label{eq:LD}
V_\mathrm{LD}(x) = \frac{\int_0^1 I(\mu) \, J_0(x \, \phi \sqrt{1-\mu^2}) \,
\mu \, \mathrm{d} \mu}{\int_0^1 I(\mu) \, \mu \, \mathrm{d} \mu}.
\end{equation}

UD diameters in J, H and K, were obtained by fitting $V_\mathrm{UD}$
(Eq.~\ref{eq:UD}) to $V_\mathrm{LD}$ (Eq.~\ref{eq:LD}) in the domain of
validity of the UD model, i.e. $V^2 \ge 0.4$. In this domain, the relative
difference between the two visibility profiles is less than
$10^{-3}$ (see Fig.~\ref{justif} for the most darkened disk of our
sample, namely M0III stars in J band).  


%
%
\section{Major characteristics of the catalog} \label{sec:charac}
\subsection{Electronic version}

An electronic version of the catalogue is available online from the Centre
de Donn\'ees astronomiques de Strasbourg (CDS). For every calibrator star,
it provides:
\begin{enumerate}
\item Identifiers: HD and HIP numbers;
\item Coordinates: ICRS 2000.0 right ascension and declination, proper
   motion, and parallax;
\item Physical properties: spectral type;
\item Photometric measurements: B and V from SIMBAD, J, H and
   Ks from 2MASS PSC, $12\,\mu$m  from IRAS;
\item Angular diameters: LD diameter and UD diameters for J, H and
  K. A single error is given for all of them, since the correction on
  this quantity due to the center to limb variation is very small;
\item Ancillary information: variability and multiplicity
   flags, variability type from Hipparcos. This category is meant to
   distinguish between stars for which no variability nor companion
   has been detected and stars for which no information is available.
\end{enumerate}

\subsection{Infrared magnitudes}

\begin{figure}
  \resizebox{\hsize}{!}{%
    \includegraphics{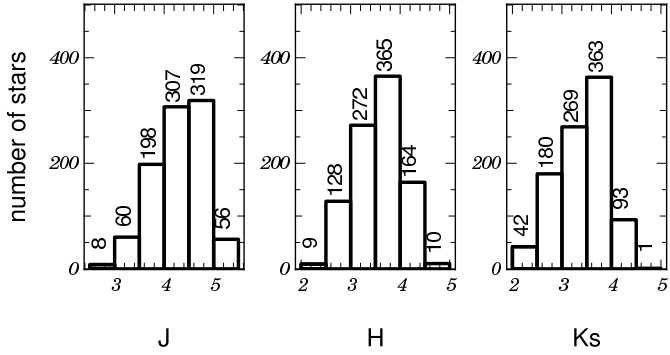}}
  \caption{Histograms of magnitudes}
  \label{histm} 
\end{figure}

  On Fig.~\ref{histm}, we plotted three histograms for J, H and Ks
  infrared magnitudes. The median magnitudes are 4.3 in J, 3.6 in H
  and 3.5 in Ks.

\subsection{Sky coverage}
\begin{figure}
   \resizebox{\hsize}{!}{%
     \includegraphics{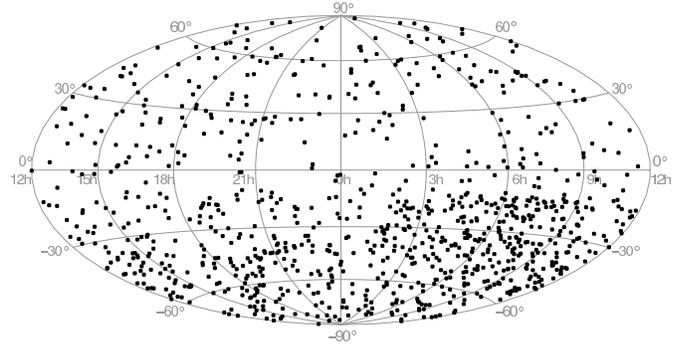}}
   \caption{Sky coverage of Cat.~2 in equatorial Hammer-Aitoff representation.}
   \label{Sky} 
\end{figure}

Cat.~2 overall sky coverage is such that there is always a calibrator
closer than $10\degr$ whatever the location on the sky
(Table~\ref{MinDist}). Moreover, if we keep only stars that can
achieve a 2\,\% visibility error for baseline greater than 200\,m in K
(145\,m in H, 110\,m in J), the calibration appears to be still feasible
for more than half the sky. The northern hemisphere happens to be less
populated than the southern one (Fig.~\ref{Sky}), because more stars
were filtered out in this part of the sky due to an unsatisfactory
spectral type identification with SIMBAD. For the southern hemisphere,
spectral identification comes from the Michigan Catalogue of Two
dimensional spectral types (\cite{Houk75}), which offers better
spectral type identification than what is available from any studies
in the northern hemisphere.

\begin{table}
   \begin{center}
     \begin{tabular}{lrr}
       \hline
       \hline
       Dist. to the closest calib. & Northern Hem. & South. Hem. \\
       \hline
       less than  3.75\degr & 45 \%  (15 \%) & 85  \% (53 \%)\\
       less than  5\degr    & 75 \%  (45 \%) & 95  \% (65 \%)\\
       less than 10\degr    & 99 \%  (85 \%) & 100 \% (89 \%)\\
       \hline
     \end{tabular}
   \end{center}
   \caption{Sky coverage characteristics: area of the sky having
     corresponding distance to the closest calibrator. Details for
     northern (positive declination) and southern (negative
     declination) hemispheres are given. The number in brackets
     corresponds to a reduced version of Cat.~2, containing only stars
     that can achieve a 2\,\% visibility error for baselines greater
     than 200\,m in K (which corresponds to 145\,m in H, 110\,m in J).}
   \label{MinDist}
\end{table}

\subsection{Baseline range}
\begin{figure}
   \resizebox{\hsize}{!}{\includegraphics{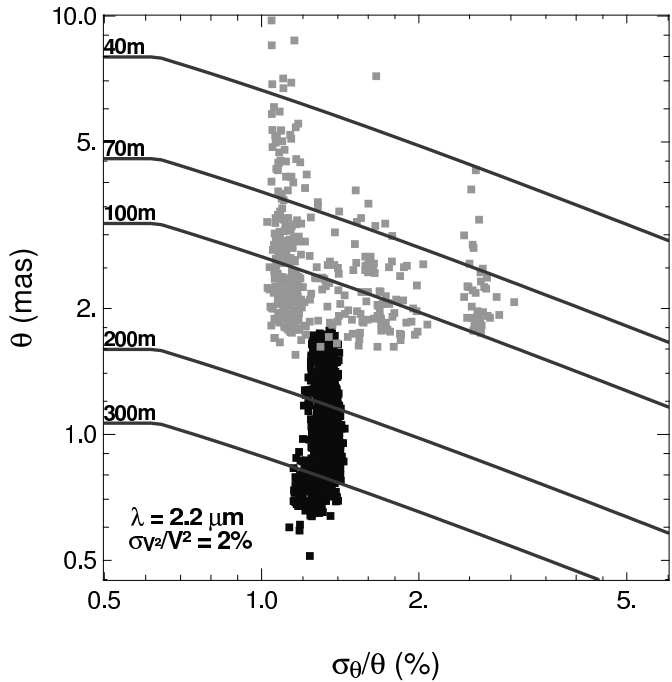}}
      \caption{Comparison of Cat.~1 (gray dots) and 2 (black dots) in
      the calibrator decision diagram. The 50, 100, 200 and 300\,m
      curves plotted here, for the K band ($\lambda=2.2\,\mu$m),
      corresponds respectively to 35, 75, 145 and 220\,m in the H band
      ($\lambda=1.6\,\mu$m) and 25, 55, 110 and 165\,m in the J band
      ($\lambda=1.2\,\mu$m).}
     \label{final_s}
\end{figure}
Fig.~\ref{final_s} shows a comparison between Cat.~1 and 2 in the
\emph{calibrator decision diagram} (observations in the K band with a
2~\% precision on the visibility), and illustrates the significant
gain due to this work. Cat.~2 can be used to achieve a 2~\% visibility
error on the whole sky with baselines up to $\sim 100$~m, $\sim
130$~m, and $\sim 180$~m, in the J, H and K bands, respectively (see
Table~\ref{MedianBase} for details). Therefore, our extended
calibrator catalog fulfills the needs for calibration of the VLTI, the
CHARA array, and any other 200-m class interferometers in the near
infrared.

\begin{table}
   \begin{center}
     \begin{tabular}{lrrr}
       \hline \hline & \multicolumn{3}{c}{$\sigma_{V^2_{\mathrm{cal.}}}
       / V^2_{\mathrm{cal.}}$} \\ Band &
       \multicolumn{1}{c}{1~\%} & \multicolumn{1}{c}{2~\%} &
       \multicolumn{1}{c}{3~\%} \\
       \hline
       J & $40 \rightarrow \:\:91$ m & $55 \rightarrow 125$ m &
       $65 \rightarrow 150$ m \\
       H & $55 \rightarrow \:121$ m & $75 \rightarrow 167$ m &
       $85 \rightarrow 200$ m \\
       K & $75 \rightarrow 163$ m & $100 \rightarrow 224$ m &
       $120 \rightarrow 266$ m \\
       
       \hline
     \end{tabular}
   \end{center}
   \caption{Improvement of the median maximum baseline for J, H and K bands
   at the 1, 2 and 3~\% visibility precision levels between Cat.~1 and 2.}
   \label{MedianBase}
\end{table}

%
%


%
%
\section{Conclusion}

We have presented a list of 1320 stars selected to serve as
calibrator stars for long baseline stellar interferometry. Among them,
374 come from the previous catalog compiled by \cite{Borde}. The
selection of the new stars was made according to the
spectro-photometric criteria defined by \cite{CohenX}, as well as
additional ones that are specific to near-infrared interferometry. We
used the work by \cite{CohenX} to estimate the angular diameter from
infrared photometric measurements extracted from IRAS, 2MASS and MSX
Astrometric catalogs. Our new catalog of interferometric calibrators
is an homogeneous set of stars, provided with limb-darkened angular
diameters and uniform disk angular diameters in J, H and K bands. Its
intrinsic characteristics -- a small list of stars with accurate
angular diameters and uniform sky coverage -- inherited from the
previous version, were carefully maintained, while we included smaller
stars in order to fulfill the needs of 200-m class interferometers,
like the VLTI in the H band (and redder) in the bright star
regime ($K \la 8$), and the CHARA array in the K band (and redder).  In
this paper, we improved the performance of Cat.~2 with respect to
Cat.~1 by extending the selection to fainter stars (this corresponds
to going downwards on the calibrator decision diagram of
Fig.~\ref{CatBorde}). As all eligible stars from the IRAS survey have
been used in Cat.~2, further extension is not possible in this
direction. Improvements could still be obtained if/when more accurate
photometry becomes available, which will result in smaller diameter
errors (this corresponds to going leftwards on the calibrator decision
diagram). For this, a dedicated observing program is needed as most of
these sources are too bright for the major infrared surveys. From the
decision diagram, it can be seen however that the gain to be expected
is in a better sky coverage at 200--300\,m baselines, rather than an
extension of the maximum baseline for which Cat.~2 would be useful.

\begin{acknowledgements}
We would like to thank the referee for his valuable corrections,
suggestions and comments. It is a pleasure to thank Stephen~T. Ridgway
and Jason~P. Aufdenberg for valuable discussions, suggestions and
corrections concerning the present work. This work has made use of the
SIMBAD database, operated at CDS, Strasbourg, France. It also has made
use of data products from the Two Micron All Sky Survey, which is a
joint project of the University of Massachusetts and the Infrared
Processing and Analysis Center/California Institute of Technology,
funded by the National Aeronautics and Space Administration and the
National Science Foundation. This work was performed in part under
contract with the Jet Propulsion Laboratory (JPL) funded by NASA
through the Michelson Fellowship Program. JPL is managed for NASA by
the California Institute of Technology.
\end{acknowledgements}



\begin{thebibliography}{}
%
\bibitem[Bessell et al. (1998)]{Bessell}
Bessell, M. S., Castelli, F., Plez, B., 1998, \aap, 333, 231

\bibitem[Bonneau et al. (2004)]{Bonneau04}
Bonneau, D., Clausse, J.-M., Delfosse, X., et al., to be published in
the proceedings of conference ``New frontiers in stellar
interferometry'' SPIE 5491, 2004

\bibitem[Bord\'e et al. (2002)]{Borde}
Bord\'e et al., 2002, \aap, 393, 183

\bibitem[Cayrel de Strobel et al. (2001)]{CdS}
Cayrel de Strobel, G., Soubiran, C., Ralite, N., 2001, \aap, 373, 159

\bibitem[Cardelli et al. (1989)]{Cardelli89}
Cardelli, J. A., Clayton, G. C., Mathis, J. S., 1989, \apj, 345, 245

\bibitem[Claret (2000)]{Claret00}
Claret, A., 2000, \aap, 363, 1081

\bibitem[C99]{CohenX}
Cohen, M., Russel, G. W., Carter, B., et al., 1999, \aj, 117, 1864 (C99)

\bibitem[Cohen et al. (2000)]{CohenXII}
Cohen, M., Hammersley, P., Egan, M., 2000, \aj, 120, 3362

\bibitem[Cohen et al. (2003)]{CohenXIV}
Cohen, M., Wheaton, Wm., A., and Megeath, S., T., 2003, \aj, 126, 1090

\bibitem[Colavita et al. 1999]{PTI}
Colavita, M.~M., Wallace, J.~K., Hines, B.~E., et al., 1999, \apj, 510, 505

\bibitem[de Medeiros and Mayor (1999)]{deMedeiros99}
de Medeiros, J.R., Mayor, M., 1999, \aaps, 139, 433 

\bibitem[Dommanget and Nys 2000]{Dommanget00}
Dommanget, J., Nys, O., 2000, \aap, 363, 991

\bibitem[Egan and Price (1996)]{Egan96}
Egan, M., Price, S., 1997, \aj, 112, 2862 

\bibitem[Glindemann et al. 2003]{vlti}
Glindemann, A., Algomedo, J., Amestica, R., et al. 2003, in
Interferometry for optical astronomy II, ed. Wesley A. Traub, Proc.  
SPIE, 4838, 89

\bibitem[Houk and Cowley 1975]{Houk75} 
Houk, N., Cowley, A.P., Michigan Catalogue of two dimentional spectral
types for the HD stars. Ann Harbor, Univ. of Michigan, 1975

\bibitem[Kervella et al. (2003)]{alphacenVINCI} 
Kervella, P., Th\'evenin, F., Morel, P., Bord\'e, P., 2003, \aap, 408, 681

\bibitem[Kervella et al. 2004]{vinci} 
Kervella, P., S\'egransan, D., Coud\'e du Foresto, V., 2004, \aap, 425, 1161

\bibitem[Leinert et al. 2004]{midi}
Leinert, C., Graser, U., Waters, L.B.F.M., et al., 2003,
in Interferometry for Optical Astronomy II. ed. Wesley A. Traub, Proc.  
SPIE, 4838, 893

\bibitem[Malaroda et al. (2001)]{Malaroda01}
Malaroda, S., Levato, H., Galliani, S., 2001, Vizier Online Data
Catalog III/216

\bibitem[Nidever et al. (2002)]{Nidever02}
Nidever, D. L., Marcy, G. W., Butler, R. P., Fischer, D. A., Vogt,
S. S., 2002, \apjs, 141, 503

\bibitem[Perrin (1998)]{Perrin98} 
Perrin , G., Coud\'e du Foresto, V., Ridgway, S.T., et al. 1998, \aap,
331, 619

\bibitem[Perryman et al. 1997]{hip}
Perryman, M. A. C., Lindegren, L., Kovalevsky, J., et al.,1999, \aap,
323, 49

\bibitem[Petrov et al. 2002]{amber} 
Petrov, R., Malbet, F., Richichi, A., et al., 2002 in Interferometry
for Optical Astronomy. ed. Pierre L\'ena and Andreas Quirrenbach, Proc
SPIE, 4006, 68

\bibitem[Quist and Lindegren 2000]{Quist2000}
Quist, C.F., Lindegren, L., 2000, \aap, 361, 770

\bibitem[Richichi et al. 2004]{charm}
Richichi, A., Percheron, I., Khristoforova, M., 2004, \aap, in press

\bibitem[ten Brummelaar et al. 2003]{chara}
ten Brummelaar, T.~A., McAlister, H.~A., Ridgway, S.~T., et al., 2003,
in Interferometry for Optical Astronomy II. ed. Wesley A. Traub, Proc.  
SPIE, 4838, 69

\bibitem[Wittkowski et al. (2004)]{Wittkowski2004}
Wittkowsky, M., Aufdenberg, J. P., Kervella, P., 2004, \aap 413, 711
\end{thebibliography}
\end{document}